\documentclass[showpacs,preprintnumbers,amsmath,amssymb,11pt]{article}
\usepackage[utf8]{inputenc}
\usepackage[english]{babel}
\usepackage{amsmath}
\usepackage{graphicx}
\usepackage[colorinlistoftodos]{todonotes}
\usepackage{caption}
\usepackage{subcaption}
\usepackage{tabularx}
\usepackage{bbm}
\usepackage[normalem]{ulem}
\usepackage{listings}
\usepackage{amsthm}
\usepackage{algorithm}
\usepackage{algpseudocode}
\usepackage{mathtools}
\usepackage{amsmath}
\usepackage[toc,page]{appendix}
\usepackage{booktabs}
\usepackage{xcolor}
\usepackage{amsfonts}
\usepackage{epstopdf}
\usepackage{adjustbox}
\usepackage{csquotes}
\usepackage{enumitem} 
\usepackage{makecell}
\usepackage{tikz}

\usetikzlibrary{decorations.pathreplacing}

\usetikzlibrary{arrows}
\usetikzlibrary{shapes}

\usepackage[numbers]{natbib}

\DeclareMathOperator*{\argmin}{arg\,min}

\makeatletter
\def\BState{\State\hskip-\ALG@thistlm}
\makeatother

\definecolor{darkgreen}{rgb}{0,0.5,0}
\definecolor{darkred}{rgb}{0.5,0,0}

\usepackage{amsmath,amsfonts,amssymb,amsthm,epsfig,epstopdf,titling,url,array}

\theoremstyle{plain}
\newtheorem{thm}{Theorem}[section]

\newtheorem{prop}[thm]{Proposition}

\theoremstyle{plain}
\newtheorem{defn}{Definition}[section]

\theoremstyle{remark}
\newtheorem{rem}{Remark}

\newcommand{\E}{\mathbb{E}}

\newcommand{\R}{\mathbb{R}}
\newcommand{\Q}{\mathbb{Q}}
\newcommand{\N}{\mathbb{N}}

\newcommand{\sd}{\textit{sd}}

\newcommand{\eqd}{\stackrel{d}{=}}

\newcommand{\arem}{$a$-remainder}
\newcommand{\VGpp}{VG++}

\newcommand{\pol}{\overline{\mathfrak{B}}}

\newcommand{\BM}{\emph{BM}}

\newcommand{\Levy}{L\'{e}vy}

\newcommand{\zapp}{Z_a^{++}}
\def\ito{It\={o}}

\title{Exchange option pricing under variance gamma-like models
}

\author{
	Matteo Gardini\thanks{Department of Mathematics, University of Genoa, Via Dodecaneso 16146, Genoa, Italy, email gardini@dima.unige.it} 	\and
	Piergiacomo Sabino\thanks{Quantitative Risk Management, E.ON SE,
		Br\"usseler Platz 1, 45131 Essen, Germany, email piergiacomo.sabino@eon.com}
		\thanks{Department of Mathematics and Statistics, University of Helsinki, P.O. Box 68 FI-00014 Finland, email piergiacomo.sabino@helsinki.fi}}

\date{\today}
\begin{document}
\maketitle

\begin{abstract}
\noindent In this article we focus on the pricing of exchange options when the dynamic of log-prices follows either the well-known variance gamma or the recent variance gamma++ process introduced in \citet{Gardini2022}. In particular, for the former model we can derive a Margrabe's type formula  whereas, for the latter one we can write  an \enquote{integral free} formula. Furthermore, we show how to construct a general multidimensional versions of the variance gamma++ processes preserving both the mathematical and numerical tractability. 

\noindent Finally we apply the derived models to German and French energy power markets: we calibrate their parameters using real market data and we accordingly evaluate exchange options with the derived closed formulas, Fourier based methods and Monte Carlo techniques. 
\end{abstract}


\section{Introduction}
Spread options are widely used in many fields of finance, in particular in energy markets, and their payoff depends on the difference of the value of two risky underlying  assets instead of one. The spread might be between spot and futures prices, between currencies, interest rates or commodities, among others. As mentioned, this type of options is particularly frequent in energy markets and the name changes depending on the commodity one is dealing with. The most notable examples are the crack, crush, and spark spreads, which measure profits in the oil, soybean, and gas markets, respectively. The interested reader can refer to \citet{Carmona2003} for a detailed discussion on this topic. By denoting $S_{i} = \left\{S_{i}(t); t \ge 0\right\}, i=1, 2$, the evolution of the value of two risky assets respectively, and labelling the strike price with $K$, the payoff of a spread option at maturity $T$ is given by
\begin{equation}
	\Phi(T) = \left(S_{2}(T) - S_{1}(T) - K\right)^{+}.
	\label{eqn:spread_option_payoff}
\end{equation}

\par The spread option valuation has been widely investigated: the first attempt to properly compute the fair price of such a contract can be found in \citet{Mar78} where the author finds a closed formula assuming a \citet{BLS1973} market and $K=0$. Unfortunately, so far, no closed form for the more general case $K \neq 0$ is known. Nevertheless, many approximate formulas have been proposed over the years. One which is commonly used has been proposed by \citet{Kirk1995} and only provides a good approximation for small values of $K$. Many other alternatives are available in the literature under the assumption of a the Black-Scholes (BS) market, see for instance \citet{Bjerksund2006} and \citet{Carmona2003}. They can be also extended taking a multi-asset spread (see for instance \citet{DLZ2008}, \citet{PellegrinoSabino_1, PellegrinoSabino_2}) or a non-Gaussian market (see for instance \citet{Pellegrino2016}). On the other hand, it is well know that price dynamics present jumps, fat tails, volatility clusters or even stochastic volatility and many other stylized facts which cannot be properly modeled by Gaussian processes. One approach to address some of the issues mentioned above is to rely on \Levy\ processes which have gained popularity because of their mathematical tractability also from the numerical point of view (see \citet{Sato} and \citet{applebaum2009} for an extensive discussion and \citet{CT2003} for applications to financial markets). On the order hand, if one leaves the Gaussian framework things might be hard to handle. Nevertheless, in some cases, such as the variance gamma (VG) process (\citet{MadanSeneta90}), the jump diffusion models proposed by \citet{Merton76} and \citet{Kou2002} and the variance gamma++ (\VGpp) process recently investigated by \citet{Gardini2022}, closed formula expressions for European options valuation can be derived. For more general models or complicated derivative structures one has to resort to numerical methods, for instance based on Monte Carlo simulations (\citet{Glass2004}), PDE's  (\citet{seydel2004} and \citet{CT2003}), Fourier methods (see \citet{Carr99}, \citet{lewis}, \citet{Attari2004}, \citet{LordFang2018}), or on the knowledge of the expression of the characteristic function (see \citet{Fang2010}). Therefore, the toolkit is rich to properly model single asset dynamics. Unfortunately, this is not always the case for the multidimensional framework. Modeling, calibration and pricing with multivariate \Levy\ processes are challenging tasks, several attempts are available in the literature: one example is the copula approach  proposed by \citet{CLV2013} and \citet{CT2003} which sometimes is difficult to handle. Moreover, in a multidimensional framework, numerical methods based on PDE or on the Fourier transform gain computational complexity. Finally, the calibration step, which is essential, might be very unstable. To address these issues \citet{Schoutens03}, \citet{Semeraro2008}, \citet{SL2010} and \citet{BB2013} proposed a framework, based on multivariate Brownian subordination, which leads to a tractable class of multidimensional \Levy\ models both from a numerical and a theoretical point of view. However, in such frameworks, closed formula expressions for spread options are rare (see for instance \citet{NCPPS2018}) and numerical methods, such as those proposed by \citet{Choi2018}, \citet{hurd2009},\citet{VanBelle2018} and \citet{caldanafusai2016}, should be used. Nevertheless, in some very particular cases, closed form expressions for very simple exchange options can be obtained giving a computational advantage and a useful benchmark to test numerical algorithms. In addition, they can be used in numerical inversion  to guess the so called \enquote{implied parameters} as observed by \citet{Carmona2003}. 
\par The first contribution of this article is the derivation of an analytic expression for the exchange options under a  bi-dimensional version of the VG model which improves the results of \citet{Hurlimann2013} and \citet{VanBelle2018}. The second contribution is the finding of an integral free Margrabe's style formula for the \VGpp\ model introduced by \citet{Gardini2022} which can be considered an extension of the VG model. To this end, we derive a multidimensional versions of the \VGpp\ process, we briefly study its properties, we provide closed form expressions of their characteristic function and that of the linear correlation coefficient. Accordingly, we compare the performance of the new formulas.

\par The article is organized as follows: in Sections \ref{sec:MargrabeFormula} and \ref{sec:MargrabeVG} we recall the Margrabe formula and extend it to the VG setting. In Section \ref{sec:VGpp_process} we recall the notion of self-decomposability and we briefly introduce the \VGpp\ process. These concepts are preparatory for the derivation  of the integral free pricing formula for exchange options in the \VGpp\ framework presented in Section \ref{sec:MargrabeVGpp}. Then, in Section \ref{sec:MultidimensionalModeling} we derive multidimensional versions of the \VGpp\ process, we briefly study its properties, provide closed form expressions of their characteristic function and that of the linear correlation coefficient. Section \ref{sec:NumericalResults} contains the applications to concrete problems: we consider electricity markets and calibrate the model parameters to real  data and price exchange options written on futures power German and France calendar contracts using the new closed formulas, Monte Carlo methods and Fourier techniques. Section \ref{sec:conclusions} concludes the paper and discusses possible extensions along with further inquires. 

\section{The Margrabe's formula}
\label{sec:MargrabeFormula}
In this section we briefly recall some well known results which are preparatory for the sequel. Consider a market with two risky assets, $S_{1}$ and $S_{2}$ and the money market account $M$, with the following dynamics under the risk neutral measure $\mathbb{Q}$

\begin{equation}
	\begin{split}
		dM(t) =rM(t)dt, \quad dS_{i}(t) = \mu_{i}S_{i}(t)dt + \sigma_{i}S_{i}(t)dW_{i}^{\Q}(t),\quad i=1,2,\\
	\end{split}
	\label{eqn:stock_dynamic}
\end{equation}
where $W_{1}^{\Q}$ and $W_{2}^{\Q}$ are  Brownian motions such that $\E\left[W_{1}^{\Q}(t)W_{2}^{\Q}(t)\right]=\rho t$. Consider the pricing problem of an exchange option, whose payoff at maturity $T$ is given by
\begin{equation}
	\Phi(T) = \left(S_{2}(T) - S_{1}(T)\right)^{+},
	\label{eqn:eachange_payoff}
\end{equation}
namely a spread option with a zero strike price. By using a change of numéraire technique, presented in \citet{Shreve04}, it is easy to show that the value of this contract $V(0)$ at time $t=0$ is given by
\begin{equation}
			V(0) = e^{(\mu_{2}-r)T}S_{2}(0)\mathcal{N}(d_{1}) - e^{(\mu_{1}-r)T}S_{1}(0)\mathcal{N}(d_{2}),
	\label{eqn:GeneralMargrabe}
\end{equation}
where
\begin{equation*}
	d_1 = \frac{\log\left(\frac{S_{2}(0)}{S_{1}(0)}\right) + \frac{1}{2}\bar{\sigma}^{2}T}{\bar{\sigma}\sqrt{T}},\quad d_{2} = d_{1} - \bar{\sigma}\sqrt{T}, \quad \bar{\sigma}^{2} = \sigma_{1}^{2} + \sigma_{2}^{2} - 2\rho\sigma_{1}\sigma_{2}.
\end{equation*}
Note that if $\mu_{1}=\mu_{2}=r$ one gets the well known formula which was derived by \citet{Mar78}. If the value of the strike price $K$ is different from zero the payoff at time $T$ is given by Equation \eqref{eqn:spread_option_payoff}  and no closed pricing formula is known therefore, numerical techniques or approximations must be adopted to properly evaluate the derivative. \\
\par It is customary to model the asset prices by the so called \Levy\ exponentiation framework, presented in \citet{CT2003}. The idea is to model the asset price process $S=\left\{S(t); t \ge 0\right\}$ as $S(t) = e^{rt + X(t)}$, where $X=\left\{X(t); t \ge 0\right\}$ is a \Levy\ Process. If $X$ is a Brownian motion with drift, we talk about Brownian exponentiation and we reduce to the original BS framework. In particular,  for $T>0$, we assume that
\begin{align*}
	M(T) = M(0)\exp\left(rT\right),\quad
	S_{i}(T) = S_{i}(0)\exp\left(\omega_{i}T + rT + \theta_{i}T + \sigma_{i}W_{i}^{\Q}(T)\right), \quad i=1,2,\\
\end{align*}
where $\E\left[W_{1}^{\Q}(t)W_{2}^{\Q}(t)\right] = \rho t$ and $\omega_{1}$ and $\omega_{2}$ are the drift corrector parameters\footnote{In BS model we have that $\omega_{i}= -\left(\theta_{i} +\frac{\sigma_{i}^{2}}{2}\right)$, for $i=1,2$.}: they are chosen so that the discounted price processes are martingales.
Applying \ito's Lemma we obtain the dynamic of $S_{1}$ and $S_{2}$ which is given by
\begin{align*}
	\frac{dS_{i}(t)}{S_{i}(t)} &= \left(r + \omega_{i} + \theta_{i} + \frac{\sigma_{i}^{2}}{2}\right)dt + \sigma_{i}dW_i^{\Q}(t),\quad i=1,2.
\end{align*}
In order to price exchange options we can use Equation \eqref{eqn:GeneralMargrabe} where we take:
\begin{equation*}
	\mu_{i}  = r + \omega_{i} + \theta_{i} + \frac{\sigma_{i}^{2}}{2}, \quad i=1,2.
\end{equation*}
This formula allows us to easily cope with the more general problem of pricing an exchange option under a time changed stochastic process like, for example, the VG model introduced by \citet{MadanSeneta90}, which overcomes some limits of the standard BS approach.

\section{Margrabe's formula under the VG model}
\label{sec:MargrabeVG}
In this section, we derive a Margrabe style option pricing formula assuming that the asset log-price dynamic follows a VG process. Let $G = \left\{G(t);t\ge 0\right\}$ be a gamma subordinator with law $\Gamma\left(\alpha t, \beta \right)$, we model the evolution of the price of the risky assets under the risk neutral measure as follows
\begin{equation*}
	S_{i}(T) = e^{\omega_{i}T + rT + \theta_{i}G(T) + \sigma_{i}W_{i}(G(T))}, \quad i=1,2,
\end{equation*}
where $\E\left[W_{1}(t)W_{2}(t)\right] = \rho t$ and:
\begin{equation*}
\omega_{i} = \alpha t \cdot \log\left(1 - \frac{\theta_{i}}{\beta} - \frac{\sigma_{i}^{2}}{2\beta}\right).
\end{equation*}	
Note that the two dynamics share the same subordinator $G$. This choice ensures that the couple $\left(\log\left(S_{1}(t)\right), \log\left(S_{2}(t)\right)\right)$ is a \Levy\ process. If we use two different subordinators, $G_{1}$ and $G_{2}$, there is no guarantee that the couple is a \Levy\ process, as observed in \citet{Buchman2017}, \citet{Buchman2019} and \citet{Michaelsen2018}.
\begin{rem}
	In financial modeling it is customary to impose that $\E\left[G(t)\right]= t$, which means that, on average, the stochastic time runs as fast as the deterministic one. This condition entails some parameter constrains: for example, if $G(t)\sim \Gamma\left(\alpha t,\beta\right)$, we have $\E\left[G(t)\right] = \alpha t/\beta$ and hence $\alpha=\beta$. Nevertheless, we keep our setting as general as possible but adopt such an assumption in the numerical section.
\end{rem}

\par Under the usual risk-neutral argument, the price of an exchange option at time $t=0$ is
\begin{equation*}
	\begin{split}
		V(0) & = e^{-rT}\E^{\Q}\left[\left(S_{2}(T) - S_{1}(T)\right)^{+}\right] \\ & =   \E^{\Q}\left[\E^{\Q}\left[\left.S_{2}(0)e^{\omega_{2} T + \theta_{2}g + \sigma_{2}W_{2}(g)} - S_{1}(0)e^{\omega_{1} T + \theta_{1}g + \sigma_{1}W_{1}(g)}\right] \right |G(T)=g]\right] \\
		 & = \E^{\Q}\left[S_{2}(0)e^{\omega_{2}T + \left(\theta_{2} + \frac{\sigma_{2}^{2}}{2}\right)G(T)} \mathcal{N}\left(d_{1}(G(T))\right) - 
		S_{1}(0)e^{\omega_{1}T + \left(\theta_{1} + \frac{\sigma_{1}^{2}}{2}\right)G(T)} \mathcal{N}\left(d_{2}(G(T))\right) \right],
	\end{split}
\end{equation*}
where in the last step, we used the Margrabe's formula derived in Equation \eqref{eqn:GeneralMargrabe} and where

\begin{align}
	d_{1}(g) & = \frac{\log\left(\frac{S_{2}(0)}{S_{1}(0)} + \left(\omega_{2} - \omega_{2}\right)T + \left(\theta_{2} - \theta_{1}\right)g + \frac{1}{2}\left(\sigma_{2}^{2} - \sigma_{1}^{2}\right)g + \frac{1}{2}\bar{\sigma}g\right)}{\bar{\sigma}\sqrt{g}}, \\
	d_{2}(g) & = d_{1}(g) - \bar{\sigma}\sqrt{g}.
	\label{eqn:d1andd2}
\end{align}
Since the probability density function of $\Gamma\left(\alpha T,\beta\right)$ is given by
\begin{equation*}
	f\left(x;\alpha T,\beta\right) = \frac{\beta^{\alpha T}}{\Gamma\left(\alpha T\right)} x^{\alpha T -1}e^{-\beta x} \mathbbm{1}_{x >0},
\end{equation*}
we get:

\begin{equation}
	\begin{split}
		V(0) & = \underbrace{\int_{0}^{\infty} S_{2}(0)e^{\omega_{2}T + \left(\theta_{2} + \frac{\sigma_{2}^{2}}{2}\right)g} \mathcal{N}\left(d_{1}(g)\right)\frac{\beta^{\alpha T}}{\Gamma\left(\alpha T\right)} g^{\alpha T -1}e^{-\beta g}dg}_{I_{1}}  \\
		& -\underbrace{\int_{0}^{\infty}S_{1}(0)e^{\omega_{1}T + \left(\theta_{1} + \frac{\sigma_{1}^{2}}{2}\right)g} \mathcal{N}\left(d_{2}(g)\right) \frac{\beta^{\alpha T}}{\Gamma\left(\alpha T\right)} g^{\alpha T -1}e^{-\beta g}dg}_{I_{2}}.
	\end{split}
	\label{eqn:integral_option_price}
\end{equation}
defining
\begin{equation*}
	\psi\left(a,b;\gamma\right) = \int_{0}^{\infty} \mathcal{N}\left(\frac{a}{\sqrt{u}} + b \sqrt{u}\right) \frac{u^{\gamma-1}}{\Gamma\left(\gamma\right)}e^{-u}du,
\end{equation*}
and focusing on the computation of $I_{1}$  we have that
\begin{equation*}
	\begin{split}
		I_{1} &= S_{2}(0) \beta^{\alpha T} e^{\omega_{2}T} \int_{0}^{\infty} \frac{1}{\Gamma\left(\alpha T\right)} g^{\alpha T -1} e^{-\left(\beta - \left(\theta_{2} + \frac{\sigma_{2}^{2}}{2}\right)\right)g} \\ &\mathcal{N}\left(\frac{\log\left(\frac{S_{2}(0)}{S_{1}(0)}\right) + \left(\omega_{2} - \omega_{2}\right)T}{\bar{\sigma}\sqrt{g}} + \frac{\theta_{2} - \theta_{1} + \frac{1}{2} \left(\sigma_{2}^{2} - \sigma_{1}^{2} + \bar{\sigma}^{2}\right)}{\bar{\sigma}}\sqrt{g}\right) dg.
	\end{split}
\end{equation*}
Now defining
\begin{equation*}
	A = \beta - \left(\theta_{2} + \frac{\sigma_{2}^{2}}{2}\right),\qquad u = Ag,
\end{equation*}
we get
\begin{equation*}
	I_{1} = S_{2}(0) \beta^{\alpha T} e^{\omega_{2} T} \int_{0}^{\infty} \frac{1}{\Gamma\left(\alpha T\right)} \frac{1}{A^{\alpha}} u^{\alpha -1} e^{-u} \mathcal{N}\left(\frac{\tilde{a}}{u} + \tilde{b}\sqrt{u}\right),
\end{equation*}
where
\begin{equation}
	\tilde{a} = \frac{\log\left(\frac{S_{2}(0)}{S_{1}(0)} + \left(\omega_{2} - \omega_{1}\right)T\right)\sqrt{A}}{\bar{\sigma}},\qquad
	 \tilde{b} = \frac{\theta_{2} - \theta_{1} + \frac{1}{2}\left(\sigma_{2}^{2} - \sigma_{1}^{2} + \bar{\sigma^{2}}\right)}{\bar{\sigma} \sqrt{A}}.
	 \label{eqn:atildebtilde}
\end{equation}
Finally, we obtain:
\begin{equation*}
	I_{1} = S_{2}(0)\frac{\beta^{\alpha T}e^{\omega_{2}T}}{A^{\alpha T}} \psi\left(\tilde{a}, \tilde{b}, \alpha T\right).
\end{equation*}
The second integral $I_{2}$ can be computed in a similar way. defining
\begin{equation*}
	C = \beta - \left(\theta_{1} + \frac{\sigma_{1}^{2}}{2}\right),
\end{equation*}
and
\begin{equation*}
	\tilde{c} = \frac{\log\left(\frac{S_{2}(0)}{S_{1}(0)} + \left(\omega_{2} - \omega_{1}\right)T\right)\sqrt{C}}{\bar{\sigma}},\qquad
	\tilde{d} = \frac{\theta_{2} - \theta_{1} + \frac{1}{2}\left(\sigma_{2}^{2} - \sigma_{1}^{2} + \bar{\sigma^{2}}\right)}{\bar{\sigma} \sqrt{C}},
\end{equation*}
we obtain that 
\begin{equation*}
I_{2} = S_{1}(0) \frac{\beta^{\alpha T}e^{\omega_{1}T}}{C^{\alpha T}}\psi\left(\tilde{c}, \tilde{d}, \alpha T\right).
\end{equation*}

Finally, the price of the exchange option in the VG model is given by

\begin{equation}
	V(0) = S_{2}(0) \frac{\beta^{\alpha T}e^{\omega_{1} T}}{A^{\alpha}}  \psi\left(\tilde{a},\tilde{b}; \alpha T\right) - S_{1}(0) \frac{\beta^{\alpha T}e^{\omega_{1} T}}{C^{\alpha}}  \psi\left(\tilde{c},\tilde{d}; \alpha T\right).
	\label{eqn:VG_spread_exact_formula}
\end{equation}

\par The function $\psi\left(a,b;\gamma\right)$ can be written in terms of the modified Bessel function of the second type (see \citet[pag.~374]{abramowitzstegun1964}) and of the confluent hypergeometric function of two variables introduced by \citet{humbert1922} hence, it can be efficiently computed, as shown in \citet{MadanCarrChang1998}. Under the additional hypothesis $S_{1}(0)=S_{2}(0)$ and $\mu_{1} = \mu_{2}$, \citet{VanBelle2018} have shown that a closed form option pricing formula for exchange options can be obtained in terms of the gamma and of the hypergeometric functions (see \citet[pag~556]{abramowitzstegun1964}). It is easy to check that such a result is a particular case of the more general formula presented in Equation \eqref{eqn:VG_spread_exact_formula}.

\begin{figure}
	\centering
	\includegraphics[scale=0.3]{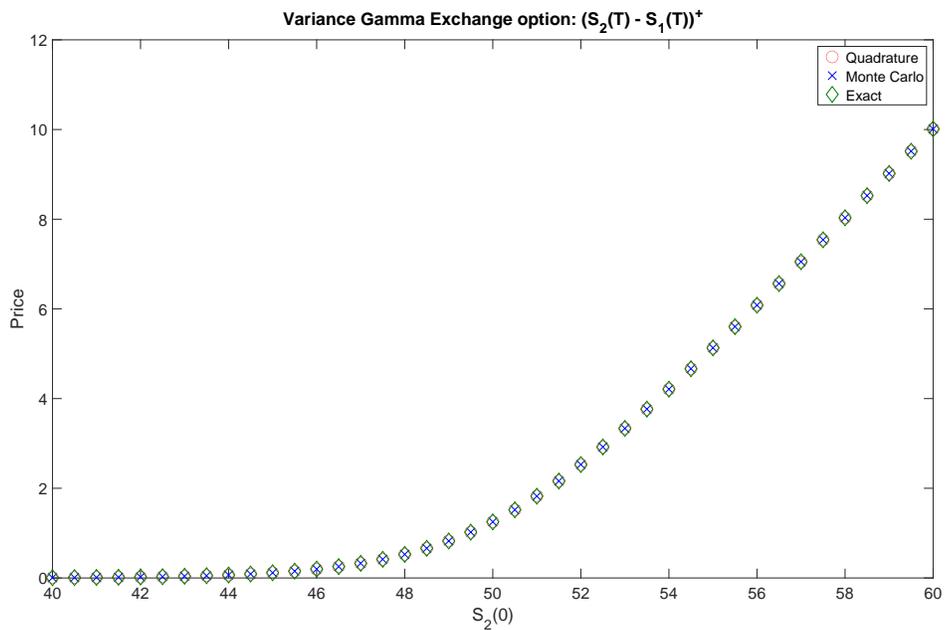}
	\caption{Price of an exchange option with $S_{1}(0)=50$ priced by using quadrature, Monte Carlo simulation and the closed formula.}
	\label{fig:VG_exhange_pricing}
\end{figure}

In Figure \ref{fig:VG_exhange_pricing} we compare three different methods to compute the exchange option price under the VG model: a standard Monte Carlo method (see \citet{CT2003}), a Gauss-Kronrod quadrature method to directly integrate Equation \eqref{eqn:integral_option_price} and finally the closed formula in Equation \eqref{eqn:VG_spread_exact_formula}. We observe that all methods produce the same results, as expected. Of course, the direct integration approach should be implemented carefully since the integrand function is quite complex. In Table \ref{tbl:speedVGpricing} we compare the computational time of both the Monte Carlo approach with $10^{6}$ simulations and the closed formula. We observe that the closed formula is approximately one hundred times faster than the Monte Carlo method and therefore such an approach is the preferable one\footnote{The computational performance has been measured using MATLAB on a PC with an Intel Core i5-10210U 2.11 GHz processor.}.

	\begin{table}
	\centering
	\small
	\begin{tabular}{cc}
		\toprule
		Algorithm & Computational time (s) \\
		\midrule
		Monte Carlo &  $2.60 \cdot 10^{-1}$  \\
		Closed formula &  $3.91 \cdot 10^{-3}$  \\
		\bottomrule
	\end{tabular}
	\caption{Computational time to price an exchange option with the closed formula and   $N_{sim}=10^{6}$ Monte Carlo simulations: the resulting time of the Monte Carlo is the average of $100$ runs.}
	\label{tbl:speedVGpricing}
\end{table}


\section{Self-decomposability and the \VGpp\ process}
\label{sec:VGpp_process}
\citet{Gardini2022} have studied a new \Levy\ process  called the \VGpp\ process. Its construction relies upon the notion of self-decomposability which on the other hand can be also used to built multidimensional \Levy\ processes including stochastic delay in the propagation of market news, as discussed in  \citet{gardini2021, gardini2021a}.
\par In defining the \VGpp\ process the main idea  is to replace the gamma subordinator by another subordinator strictly related to it which in turn, has finite activity and the step-wise increasing trajectories. Therefore, seeing the subordinator as the evolution of the number of transactions, the probability of having no transaction in a finite period of time will not be in general zero. 
\par We recall that a random variable $X$ is said  to have a self-decomposable law if for all $a \in \left(0,1\right)$ we can always find two independent random variables $Y$, with the same law as $X$, and $Z_{a}$ such that in distribution we have
\begin{equation}
	X \eqd aY + Z_{a}.
	\label{eqn:sd_relation}
\end{equation}
We refer to $Z_{a}$ as the \arem\ of the \sd\ law.
Accordingly, denoting $\phi_{X}\left(u\right)$ the characteristic function  of $X$ and by $\phi_{Z_{a}}\left(u\right)$ that of $Z_{a}$ we have that
\begin{equation}
	\phi_{X}\left(u\right)= \phi_{X}\left(au\right)\phi_{Z_{a}}\left(u\right).
	\label{eqn:chfsdrelation}
\end{equation}
The law of $Z_{a}$ turns out to be infinitely divisible (see \citet[Proposition~15.5]{Sato}) and  one can construct the associated \Levy\ process $Z_{a}=\left\{Z_{a}(t);t\ge 0\right\}$ which is actually a subordinator that can be used for Brownian subordination. \\
\par The gamma distribution is a well-known self-decomposable law (see \citet{Gri03}) and the distribution $\Gamma^{++}\left(a,\alpha, \beta\right)$ of its \arem\ $Z_{a}$ is infinitely divisible. The \VGpp\ process is then defined as follows
\begin{defn}
	Consider a Brownian motion $W=\left\{W(t);t\ge0\right\}$, with drift $\theta \in \R$, diffusion $\sigma \in \R^+$ independent of $\zapp$. We call the process $X = \left\{X(t);t\ge0\right\}$  defined as
	\begin{equation}
		X(t) = \theta Z_{a}(t) + \sigma W\left(Z_{a}(t)\right), \quad t \ge 0
		\label{eqn:VG++}
	\end{equation}
	\VGpp\ process.
\end{defn}

Such a process has been studied in details in \citet{Gardini2022} and it has several nice properties. It can be written as difference of two $\zapp$ processes with suitable parameters, it is of finite activity and therefore of finite variation. In particular, the probability density function of $X$ at time $t>0$ and its characteristic function can be written in closed form. In addition, the path-simulation is very efficient. Accordingly, all the common numerical techniques for calibration and pricing can be adopted. Finally, a numerically convenient integral-free closed form option pricing formula for European call (and put) options is available. 
\par From an economical point, the \VGpp\ process can be used to model illiquid markets. Indeed, the value of the parameters $a$ and $\alpha$ are related to the liquidity activity of the market. Taking the
change $\Delta X = X(t) - X(t-1)$ of the log-price over the time interval $\Delta t$, it can be shown that the probability
that the increment equals zero over the time interval $\Delta t$ is strictly larger than zero and, more precisely, it is given by
\begin{equation*}
	P\left(\Delta X = 0\right)= a^{ \alpha \Delta t},
\end{equation*}
since the density of the VG++ process has an atom in zero. This is the main financial difference from the standard VG
process which does imply that non-zero trading activity takes place in every time interval. On the other hand, the \VGpp\ model inherits the mathematical tractability of the standard VG process which is, in any case, recovered when $a$ tends to zero. 
\par It turns out that an integral free closed formula for exchange options can obtained for the \VGpp\ model as well, which will be detailed in the next section.

\section{An integral free formula for exchange options under the VG ++ process}
\label{sec:MargrabeVGpp}
In this section we derive the explicit formula for exchange options under the \VGpp\ process but first recall some useful relations. 
\par Following \citet{humbert1922}, we define the confluent hypergeometric function of two variables $x$ and $y$ as
\begin{equation}
	\Phi\left(\alpha,\beta, \gamma; x,y\right) = \frac{\Gamma(\gamma)}{\Gamma(\alpha)\Gamma(\gamma-\alpha)}\int_{0}^{1} u^{\alpha -1} \left(1-u\right)^{\gamma-\alpha-1}\left(1-ux\right)^{-\beta}e^{uy}du,
	\label{eqn:confluent_hypergeometric}
\end{equation}
and observe that if $\alpha= n, \beta = 1-n$ and $\gamma=1+n$ for $n \in \N$, Equation \eqref{eqn:confluent_hypergeometric} becomes:
\begin{equation}
	\Phi\left(n,1-n, 1+n; x,y\right) = n\int_{0}^{1} u^{n-1} \left(1-ux\right)^{n-1}e^{uy}du.
	\label{eqn:confluent_hypergeometric_integer}
\end{equation}
It turns out that the integral in Equation \eqref{eqn:confluent_hypergeometric_integer} can be computed exactly, as it will be shown later. $\Phi\left(1+n,1-n, 2+n; x,y\right)$ can be computed in a similar way.
\par Another important formula relative to the Bessel function of the second kind $K_{\nu}$ when $\nu= n + \frac{1}{2}$ (see \citet[pag.~444]{abramowitzstegun1964}) is

\begin{equation}
	\sqrt{\frac{\pi}{2x}} K_{n+\frac{1}{2}}(x) = \left(\frac{\pi}{2}\right)e^{-x}\sum_{k=0}^{n}\left(n+\frac{1}{2},k\right)\left(2x\right)^{-k},
	\label{eqn:besselk_integer}
\end{equation}
where
\begin{equation*}
	\left(n+\frac{1}{2},k\right) = \frac{\left(n+k\right)!}{k! \Gamma\left(n-k+1\right)}.
\end{equation*}
Finally, the following trivial identity will be used
\begin{equation*}
	K_{n-\frac{1}{2}}(x) = K_{(n-1)+\frac{1}{2}}(x),\quad n \in \N.
\end{equation*}

\subsection{Computation of $\Phi\left(n,1-n, 1+n; x,y\right)$}
\label{sec:explicit_expressions}
There are several possible ways to compute Equation \eqref{eqn:confluent_hypergeometric_integer} analytically. Observing that

\begin{equation}
	\begin{split}
		I_{2} & = \int_{0}^{1} \left(u(1-ux)\right)^{n-1}e^{uy}du = \int_{0}^{1} \left(u(1-ux)\right)^{n-1}\sum_{m=0}^{\infty}\frac{(uy)^{m}}{m!} du \\
		& = \sum_{m=0}^{\infty} \frac{y^{m}}{m!} \int_{0}^{1} u^{m+n-1}\left(1-ux\right)^{n-1}du,
	\end{split}
\end{equation}
the problem reduces to compute an integral of the form

\begin{equation*}
	I_{3} = \int u^{p}\left(1-ux\right)^{q}du = I_{3,1} + I_{3,2} + I_{3,3} + C.
\end{equation*}
Simple computations show that
\begin{align*}
	I_{3,1} & = \frac{u^{p+1}}{p+1}\left(1-ux\right)^{q}, \\
	I_{3,2} & = \sum_{j=2}^{q} x^{j-1}\left(1-ux\right)^{q+1-j} \frac{\prod_{k=2}^{j} (q+2-k)}{\prod_{k=1}^{j}(p+j)},\\
	I_{3,3} & = x^{q}q! \frac{u^{p+q+1}}{(p+1)(p+2)\dots(p+q)(p+q+1)}.
\end{align*}
and hence we obtain an analytical expression for Equation $\Phi\left(n,1-n, 1+n; x,y\right)$.\\
\par On the other hand, relying upon the binomial theorem
\begin{equation*}
	\left(x+y\right)^{n} = \sum_{k=0}^{n} \binom{n}{k} x^{n-k}y^{k},
\end{equation*}
it follows that

\begin{equation}
	\begin{split}
		I_{2} &= \int \left(u-u^{2}x\right)^{n-1}e^{uy}du = \sum_{k=0}^{n-1}\binom{n-1}{k} \int u^{n-1-k}\left(-u^{2}x\right)^{k}e^{uy}du \\
		& = \sum_{k=0}^{n-1}\left(-1\right)^{k}x^{k}\binom{n-1}{k}\underbrace{\int u^{n-1+k}e^{uy}du}_{I_4}.
	\end{split}
\end{equation}
Since:

\begin{equation*}
	\int u^{m}e^{uy} du = e^{uy}\cdot \left\{\sum_{k=0}^{m} \left(-1\right)^{k} \binom{m}{k}k!\frac{u^{m-k}}{y^{k+1}}\right\},
\end{equation*}
we can conclude that

\begin{equation*}
	\Phi\left(n,1-n, 1+n; x,y\right) = n \sum_{k=0}^{n-1}\left\{\left(-1\right)^{k} x^{k}\binom{n-1}{k} \cdot e^{uy}\cdot \left(\sum_{j=0}^{m} \left(-1\right)^{j}\binom{m}{j}j\frac{u^{m-j}}{y^{j+1}}\right)\right\}.
\end{equation*}

\subsection{Exchange option Pricing}
In this section we derive a Margrabe's style formula for exchange option for the \VGpp\ model. 
For $T>0$, we model the two risky asset $S_{1}$,$S_{2}$ under measure $\Q$ as follows
\begin{equation}
	\begin{split}
	S_{i}(T) = S_{i}(0) e^{\omega_{i}T + rT + \theta_{i}Z_{a}(T) + \sigma_{i}W_{i}\left(Z_{a}(T)\right)},\quad i=1,2,
\end{split}
\label{eqn:VGpp_undelyings_Schoutens}
\end{equation}
where $W_{1}=\left\{W_{1}(t); t\ge 0\right\}$ and $W_{2}=\left\{W_{2}(t); t\ge 0\right\}$ are two standard Brownian motions, $\E\left[W_{1}(t)W_{2}(t)\right] = \rho t$ and $Z_{a}=\left\{Z_{a}(t); t \ge 0\right\}$ is a $\Gamma^{++}\left(a,\alpha t, \beta \right)$ subordinator, as discussed in Section \ref{sec:VGpp_process}. Observe that we use the subordinator $Z_{a}$ to time change both Brownian motions $W_{1}$ and $W_{2}$ as proposed by \citet{Schoutens03}: this guarantees us to remain in a \Levy\ framework also when $W_{1}$ and $W_{2}$ are correlated. From \citet{Gardini2022} we recall that the density of $Z_{a}\sim \Gamma^{++}\left(a,\alpha, \beta\right)$ is given by

\begin{equation*}\
	h_{a}(x) = a^{\alpha}\delta_{0}(x) + \sum_{n\ge 1} \binom{\alpha + n -1}{n}a^{\alpha} \left(1-a\right)^{n} f_{n,\beta/a}(x) \mathbbm{1}_{x\ge 0},
\end{equation*}
where, for $\alpha >0$,
\begin{equation*}
	\binom{\alpha + n -1}{n} = \frac{\alpha \left(\alpha-1\right)\dots \left(\alpha - k +1\right)}{k!}, \quad \binom{\alpha}{0} = 1,
\end{equation*}
and
\begin{equation*}
	f_{n,\beta/a} = \left(\frac{\beta}{a}\right)^{n} x^{n-1} \frac{e^{-\beta x/a}}{\Gamma\left(n\right)},
\end{equation*}
which is the density of an Erlang distribution with parameters $n \in \N$ and $\beta/a$.
Proceeding by conditioning on $Z_{a}(T)$, as done in the VG case, we have that the value of the exchange option $V(0)$ at time $t=0$ is given by

\begin{equation*}
	\begin{split}
	V(0) & = e^{-r T} \E^{\mathbb{Q}}\left[\left(S_{2}(T)-S_{1}(T)\right)^{+}\right]\\
	& = \E^{\mathbb{Q}}\left[S_{2}(0)e^{\omega_{2}T + \left(\theta_{2} + \frac{\sigma_{2}^{2}}{2}\right)Z_{a}(T)}\mathcal{N}\left(d_{1}\left(Z_{a}(T)\right)\right) - S_{1}(0)e^{\omega_{1}T + \left(\theta_{1} + \frac{\sigma_{1}^{2}}{2}\right)Z_{a}(T)}\mathcal{N}\left(d_{2}\left(Z_{a}(T)\right)\right)\right],
	\end{split}
\end{equation*}
where $d_{1}(x)$ and $d_{2}(x)$ are given by Equation \eqref{eqn:d1andd2}.
\par We use the linearity of expected value and observe that we can focus only on the computation of the first expectation, since the one of the second term is similar. Omitting the the subscripts we have that
\begin{equation}
	\begin{split}
	& \E^{\mathbb{Q}}\left[S(0)e^{\omega T + \theta Z_{a}(T) + \frac{\sigma^{2}}{2}Z(T)}\mathcal{N}\left(d_{1}\left(Z_{a}(T)\right)\right)\right]  = S(0)e^{\omega T}\int_{0}^{\infty}e^{\theta x + \frac{\sigma^{2}}{2}x}\mathcal{N}\left(d_{1}(x)\right)h_{a}(x)dx \\
	& = \psi \cdot a^{\alpha T} S(0)e^{\omega T} + S(0)e^{\omega T} \sum_{n\ge 1} \binom{\alpha T + n -1}{n}a^{\alpha T}\left(1-a\right)^{n} \underbrace{\int_{0}^{\infty} e^{\theta x + \frac{\sigma^{2}}{2}x} f_{n,\beta/a}(x)\mathcal{N}\left(d_{1}(x)\right)dx}_{I_{1}},
	\end{split}
\label{eqn:exact_formula}
\end{equation}
where

	\[ \psi =  \begin{cases} 
		1 & \text{if } \log\left(S_{2}(0)/S_{1}(0)\right) + \left(\omega_{2} - \omega_{1} \right)T > 0, \\
		0 & \text{otherwise.}
	\end{cases}
	\]
Focusing on the computation of the term $I_{1}$
\begin{equation*}
	I_{1} = \int_{0}^{\infty} \left(\frac{\beta}{a}\right)^{n} \frac{1}{\Gamma(n)}e^{-\left(\beta/a -\left(\theta + \frac{\sigma^{2}}{2}\right)x\right)} x^{n-1}\mathcal{N}\left(d_{1}(x)\right)dx,
\end{equation*}
and defining $A_{j} = \frac{\beta}{a} - \left(\theta + \frac{\sigma^{2}}{2}\right)$ and $u = A_{j}x$ we have that
\begin{equation*}
	I_{1} = \left(\frac{\beta}{a A_{j}}\right)^{n}\int_{0}^{\infty} \mathcal{N}\left(\frac{\tilde{a}}{u} + \tilde{b}\sqrt{u}\right) \frac{u^{n-1}}{\Gamma(n)}e^{-u}du = \left(\frac{\beta}{a A_{j}}\right)^{n}\Psi\left(\tilde{a}, \tilde{b};n\right).
\end{equation*}
for properly defined values of $\tilde{a}$ and $\tilde{b}$, given by Equations \eqref{eqn:atildebtilde}. 

In particular, as shown in \citet{MadanCarrChang1998}, $\Psi\left(\tilde{a}, \tilde{b};n\right)$ can be expressed in terms of the modified Bessel function of the second kind $K_{\nu}(z)$ and $\Phi\left(\alpha, \beta, \gamma;x,y\right)$  where $\alpha$, $\beta$ and $\gamma$ assume integer values. $\Psi\left(a,b;\gamma\right)$ has the following expression

\begin{equation*}
	\begin{split}
		\Psi\left(a,b;\gamma\right) & = \frac{c^{\gamma +1/2} \exp\left[sign\left(a\right)c\right]\left(1+u\right)^{\gamma}}{\sqrt{2\pi}\Gamma\left(\gamma\right)\gamma}\cdot K_{\gamma+1/2}\left(c\right) \\
		&\Phi\left(\gamma,1-\gamma,1+\gamma,\frac{1+u}{2},-sign\left(a\right)c\left(1+u\right)\right) \\
		& -sign\left(a\right) \frac{c^{\gamma +1/2} \exp\left[sign\left(a\right)c\right]\left(1+u\right)^{\gamma+1}}{\sqrt{2\pi}\Gamma\left(\gamma\right)\left(1+\gamma\right)}\cdot K_{\gamma-1/2}\left(c\right) \\  &\Phi\left(1+\gamma,1-\gamma,2+\gamma,\frac{1+u}{2},-sign\left(a\right)c\left(1+u\right)\right) \\
		&+sign\left(a\right)\frac{c^{\gamma +1/2} \exp\left[sign\left(a\right)c\right]\left(1+u\right)^{\gamma}}{\sqrt{2\pi}\Gamma\left(\gamma\right)\gamma}\cdot K_{\gamma-1/2}\left(c\right) \\
		&\Phi\left(\gamma,1-\gamma,1+\gamma,\frac{1+u}{2},-sign\left(a\right)c\left(1+u\right)\right),
	\end{split}
\end{equation*} 
where $c = |a|\sqrt{2+b^2}$ and $u=\frac{b}{\sqrt{2+b^2}}$. $K_{\nu}(z)$ can be computed using Equation \eqref{eqn:besselk_integer} when $\nu = n \pm 1/2$, whereas, the confluent hypergeometric function in Equation \eqref{eqn:confluent_hypergeometric} has an explicit expression as observed in Section \ref{sec:explicit_expressions}.
Summarizing all the previous results we have the following proposition.

\begin{prop}
	\label{prop:VGppEchange_closedformula}
The price $V(0)$ of an exchange option at time $t=0$  with payoff at maturity $T$ given by $\Phi\left(T\right) = \left(S_{2}(T) - S_{1}(T)\right)^{+}$ under the dynamics in Equation  \eqref{eqn:VGpp_undelyings_Schoutens} is given by
\begin{equation*}
	\begin{split}
	V(0)  & = S_{2}(0)e^{\omega_{2} T} \left( \psi \cdot a^{\alpha T}  +  \sum_{n\ge 1} \binom{\alpha T + n -1}{n}a^{\alpha T}\left(1-a\right)^{n} \Psi\left(\tilde{a},\tilde{b};n\right) \frac{1}{A^{n}}\right)\\ 
	& - S_{1}(0)e^{\omega_{1} T} \left( \psi \cdot a^{\alpha T}  +  \sum_{n\ge 1} \binom{\alpha T + n -1}{n}a^{\alpha T}\left(1-a\right)^{n} \Psi\left(\tilde{c},\tilde{d};n\right) \frac{1}{C^{n}}\right),
\end{split}
\end{equation*}
where
\begin{equation*}
	A  = \frac{a}{\beta} \frac{1}{A_{2}}, \qquad C = \frac{a}{\beta} \frac{1}{A_{1}}, \qquad \psi =  \begin{cases} 
		1 & \text{if } \log\left(S_{2}(0)/S_{1}(0)\right) + \left(\omega_{2} - \omega_{1} \right)T > 0, \\
		0 & \text{otherwise.}
	\end{cases}
\end{equation*}
and
\begin{align*}
	&\tilde{a}  = \frac{\log\left(\frac{S_{2}(0)}{S_{1}(0)} + \left(\omega_{2} - \omega_{1}\right)T\right)\sqrt{A_{2}}}{\bar{\sigma}},\qquad
&\tilde{b} = \frac{\theta_{2} - \theta_{1} + \frac{1}{2}\left(\sigma_{2}^{2} - \sigma_{1}^{2} + \bar{\sigma^{2}}\right)}{\bar{\sigma} \sqrt{A_{2}}}, \\
&\tilde{c} = \frac{\log\left(\frac{S_{2}(0)}{S_{1}(0)} + \left(\omega_{2} - \omega_{1}\right)T\right)\sqrt{A_{1}}}{\bar{\sigma}},\qquad
&\tilde{d} = \frac{\theta_{2} - \theta_{1} + \frac{1}{2}\left(\sigma_{2}^{2} - \sigma_{1}^{2} + \bar{\sigma^{2}}\right)}{\bar{\sigma} \sqrt{A_{1}}},
\end{align*}
and
\begin{equation*}
	A_{1} = \frac{\beta}{a} -\left(\theta_{1} + \frac{1}{2}\sigma_{1}^{2}\right), \quad A_{2} = \frac{\beta}{a} -\left(\theta_{2} + \frac{1}{2}\sigma_{2}^{2}\right), \quad \sigma^{2} = \sigma_{1}^{2} + \sigma_{2}^{2} - 2\rho \sigma_{1}\sigma_{2}.
\end{equation*}
\end{prop}
\begin{rem}
We stress that the pricing formula for the exchange option $V(0)$ in Proposition \ref{prop:VGppEchange_closedformula} does not require numerical integration  since $\Psi\left(x,y;n\right)$ can be computed analytically for $n\in \N$ as shown in Section \ref{sec:explicit_expressions}.
\end{rem}

\subsection{Numerical test of the option pricing formula}
In this section we compare the price of an exchange option calculated with the exact formula of the previous section to those ones obtained with  Monte Carlo and Fourier-based techniques.
\par To this end, we choose the following set of parameters: $r=0.01$, $\sigma_{1} = 0.2$, $\sigma_{2}=0.3$, $\theta_{1}=-0.2012$, $\theta_{2}=-0.1712$, $S_{1}(0)=100$, $S_{2}(0)=105$, $T=1$, $\alpha=2$, $\rho=0.8$. 
\par We focus on the bi-variate \VGpp\ process for $S_{1}$ and $S_{2}$ defined in Equation \eqref{eqn:VGpp_undelyings_Schoutens} and we define
\begin{equation*}
	Y_{i}(t) = \theta_{i}Z_{a}(T) + \sigma_{i}W_{i}(Z_a(T)),\quad i=1,2.\\
\end{equation*}
The generation of the skeleton of each component is straightforward given the simulation of $Z_{a}$ which can be easily obtained relying on the method illustrated in \citet{cs20}. Namely,
$$
	Z_{a}\stackrel{d}{=} \left\{\begin{array}{ll}\sum_{i=1}^{S} X_i, &\mbox{when $S>0$}\\
		0, &\mbox{when 
			$S=0$}\end{array}\right.$$
	where $X_i\sim \mathcal E(\beta/a)$ iid and $S \sim \pol\left({\alpha,1-a}\right)$. In particular ${Z_a}_{|S=s}\sim \Gamma (s, \beta/a)$, when $s>0$.
\par Several FFT-based methods are available in the literature, among other for instance the one proposed in \citet{hurd2009} or \citet{caldanafusai2016} or that based on the cosine expansions proposed by \citet{Pellegrino2016}. All these approaches require the characteristic function of the log-distribution of prices under the risk-neutral measure in a closed form. 
To this end, it is possible to show that the characteristic function of $\boldsymbol{Y}(t) = \left(Y_{1}(t),Y_{2}(t))\right)$ is given by
\begin{equation*}
	\phi_{\boldsymbol{Y}(t)} = \phi_{Z(t)}\left(\boldsymbol{\theta}^{T}\boldsymbol{u} + \frac{i}{2}\boldsymbol{u}^{T}\Sigma \boldsymbol{u}\right),
\end{equation*}
where $\boldsymbol{\theta} = \left[\theta_{1},\theta_{2}\right]$, $\boldsymbol{u} = \left[u_{1},u_{2}\right]$ and 
\begin{equation*}
	\Sigma= 
\begin{bmatrix}
	\sigma_{1}^{2} & \rho \sigma_{1} \sigma_{2} \\
	\rho \sigma_{1} \sigma_{2} & \sigma_{2}^2
\end{bmatrix}.
\end{equation*}

\begin{figure}
	\centering
	\includegraphics[scale=0.4]{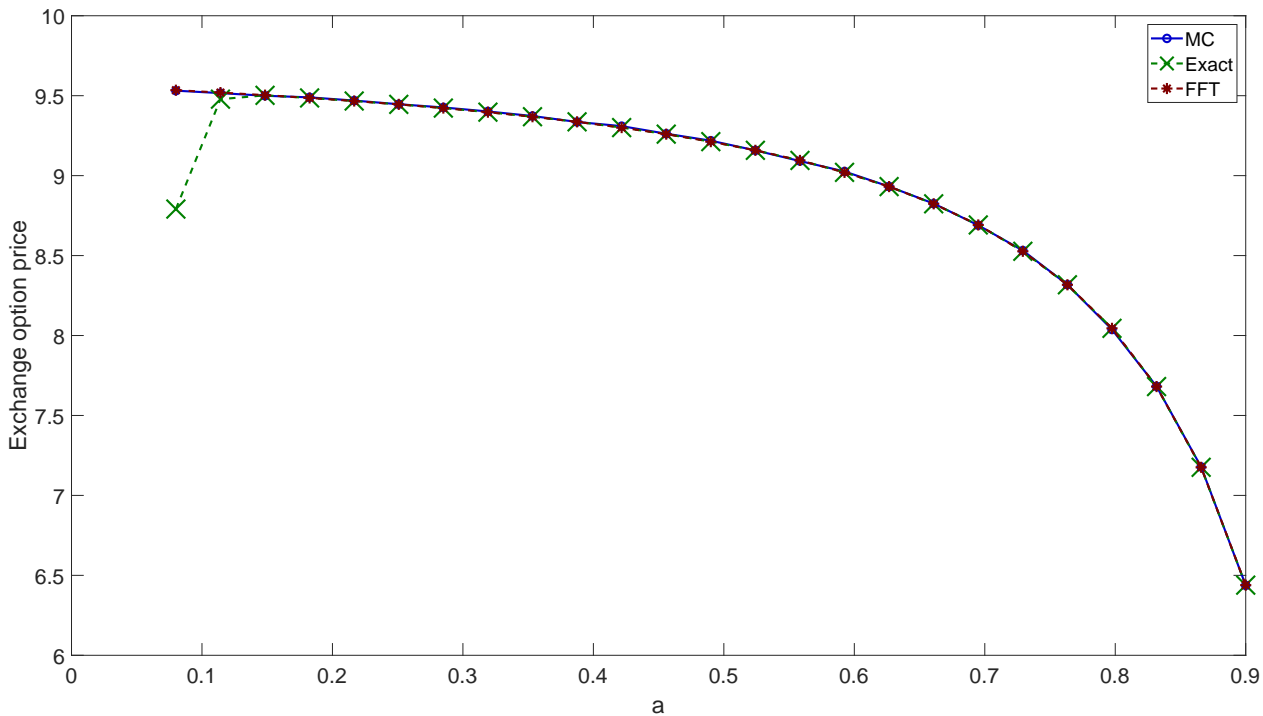}
	\caption{Exchange option pricing in the \VGpp\ model computed by exact formula, Monte Carlo simulations and FFT method.}
	\label{fig:VGppMCvsExact}
\end{figure}

The prices of the exchange options for different values of $a$ are shown in Figure \ref{fig:VGppMCvsExact}. Here we compare the analytic formula with the Monte Carlo scheme and the approximate formula for spread options proposed by \citet{caldanafusai2016}, which is simpler than the one proposed by \citet{hurd2009} since it requires a single Fourier transform inversion. In Figure \ref{fig:VGppMCvsExact} we observe that for small values of $a$ the computation of the exact formula might become problematic. This is because the exact formula requires the computation of several terms involving many factorials, which might be numerically unstable in some cases: some attention should be payed in this case to avoid  memory overflow. Furthermore, even with a smart truncation strategy, the convergence of the series in Equation \eqref{eqn:exact_formula} can be slow  and consequently the computation of the formula might require some time. For these reasons, in practical purposes, the pricing techniques based on Fourier transform or on the Monte Carlo method are preferable. Nevertheless, the exact formula provides a benchmark to test numerical algorithms and more importantly, could be inverted to infer the so called \enquote{implied parameters}, as commonly done to get the option implied volatility. 

\section{Multidimensional modeling}
\label{sec:MultidimensionalModeling}
In this section we go beyond the modeling approach used in the previous section which  relies upon a single subordinator (see \citet{LS2006}). Such  an approach can be extended  using a multidimensional subordinator while still remaining in a \Levy\ framework. Successful approaches have been proposed by \citet{Semeraro2008}, \citet{SL2010}, \citet{BB2013} and taken up by \citet{gardini2021, gardini2021a} to introduce delays in market news propagation. In order to generalize the results of Section \ref{sec:MargrabeVGpp} we recall some properties of the $\Gamma^{++}\left(a, \alpha, \beta \right)$ distribution which are directly inherited from the gamma law (see \citet{cs20}).

\begin{align*}
	&Z_{a} \sim \Gamma^{++}\left(a,\alpha, \beta\right), \quad &cZ_{a} \sim \Gamma^{++}\left(a,\alpha, \frac{\beta}{c}\right),\; c>0, \\
	&Z_{a,i} \sim \Gamma^{++}\left(a,\alpha_{i}, \beta\right), \quad 
	& \sum_{i=1}^{n}Z_{a,i} \sim \Gamma^{++}\left(a, \sum_{i=1}^{n}\alpha_{i}, \beta\right).
\end{align*}
Furthermore, the characteristic function of $Z_{a}\sim \Gamma^{++}\left(a,\alpha,\beta\right)$ is given by
\begin{equation}
	\phi_{Z_{a}}(u) = \left(\frac{\beta - iua}{\beta -iu}\right)^{\alpha}, \quad u \in \R,
	\label{eqn:chf_Zsubordinator}.
\end{equation}
Moreover, for $c \in \R^{+}$, we have that $\phi_{cZ_{a}}(u) = \phi_{Z_{a}}(cu)$.

\subsection{Semeraro's approach}
We adapt the approach illustrated in \citet{Semeraro2008}, in the context of the \VGpp process. We consider independent subordinators $I_{j} = \left\{I_{j}(t); t \ge 0\right\},\; j=1,\dots,n$ also independent from $Z_a = \left\{Z_{a}(t); t \ge 0\right\}$ and, for $\alpha_{j}\ge 0,\; j=1,\dots,n$, we consider $G_{j} = \left\{G_{j}(t); t \ge 0\right\}$ defined as

\begin{equation}
	G_{j}(t) = I_{j}(t) + \alpha_{j}Z_a(t), \quad j=1,\dots,n.
\end{equation}
By choosing:
\begin{equation}
	\begin{split}
	I_{j}(t) & \sim \Gamma^{++}\left(a, A_{j} t, \frac{B}{\alpha_{j}}\right),\quad A_{j},B >0, \\
	Z_{a}(t) & \sim \Gamma^{++}\left(a, A t, B\right), \quad A,B >0,
	\end{split}
\label{eqn:SemeraroSubDistro}
\end{equation}
we have that

\begin{equation*}
	G_{j}(t) \sim \Gamma^{++}\left(a, \left(A_{j}+A\right)t,\tfrac{B}{\alpha_{j}}\right).
\end{equation*}

Then, given a standard Brownian motion with drift $\boldsymbol{\mu}=\left(\mu_{1},\dots,\mu_{n}\right)$ and volatility $\boldsymbol{\sigma}=\left(\sigma_{1},\dots,\sigma_{n}\right)$  we can define a subordinated Brownian motion $\boldsymbol{Y} = \left\{\left(Y_{1}(t),\dots,Y_{n}(t)\right),t \ge 0\right\}$ as

\begin{equation*}
	Y_{j}(t) = \mu_{j}G_{j}(t) + \sigma_{j}W_{j}\left(G_{j}(t)\right),\quad j=1,\dots,n.
\end{equation*}
The characteristic function of $\boldsymbol{Y}$ at time $t$ can be easily computed and is given by
\begin{equation*}
	\E\left[e^{i \langle \boldsymbol{u},\boldsymbol{Y} \rangle}\right] = \prod_{j=1}^{n} \phi_{I_{j}(t)}\left(u_{j}\mu_{j} + i \frac{\sigma_{j}^{2}}{2}u_{j}^{2}\right) \cdot \phi_{Z_a(t)}\left(\alpha_{j}\left(u_{j}\mu_{j} + i \frac{\sigma_{j}^{2}}{2}u_{j}^{2}\right)\right),
\end{equation*}
whereas, the linear correlation coefficient at time $t$ has the following form

\begin{equation}
	\rho_{Y_{i}(t),Y_{i}(t)} = \frac{\mu_{i}\mu_{j}\alpha_{i}, \alpha_{j} Var\left[Z_a(t)\right]}{\sqrt{Var\left[Y_{i}(t)\right] Var\left[Y_{j}(t)\right]}}.
\end{equation}

Linear correlation coefficient is important for the model calibration while instead, the knowledge of the characteristic function allows us to use pricing methods based on Fourier techniques.
It is worth noting that the model we have just proposed is not always able to properly catch  the correlation observed in the market, especially if the drift $\mu_{j}$ is small. For this reason, \citet{SL2010} introduced an extended version of this model which aims at better fitting  the market correlation even if the drift $\mu_{j}$ vanishes.

\subsection{Luciano-Semeraro's approach}
In this section we adapt the approach of \citet{SL2010} which in contrast, introduces correlated Brownian motions (hereafter labeled LS). Consider mutually independent subordinators $I_{i} = \left\{I_{i}(t); t \ge 0\right\},\; i=1,\dots,n$ and another independent subordinator $Z_a = \left\{Z_a(t); t \ge 0\right\}$. Define the multidimensional process $\boldsymbol{Y}=\left\{Y_{1},\dots,Y_{n}\right\}$ as follows

	\begin{equation}
	\boldsymbol{Y}\left(t\right) =
	\left(
	\begin{array}{ll}
		\mu_{1} I_{1}\left(t\right) + \sigma_{1}W_{1}\left(I_{1}\left(t\right)\right) +\alpha_{1}\mu_{1}Z_{a}\left(t\right) + \sqrt{\alpha_{1}}\sigma_{1}W_{1}^{\rho}\left(Z_{a}\left(t\right)\right) \\
		\dots\\
		\mu_{n} I_{n}\left(t\right) + \sigma_{n}W_{n}\left(I_{n}\left(t\right)\right) +\alpha_{n}\mu_{n}Z_a\left(t\right) + \sqrt{\alpha_{n}}\sigma_{n}W_{n}^{\rho}\left(Z_a\left(t\right)\right)
	\end{array}\right),
	\label{eqn:LSgeneralization}
\end{equation}
where $I_{j}(t)\; j=1,\dots,n$ and $Z_a(t)$ are distributed as in Equation \eqref{eqn:SemeraroSubDistro}.
As shown by \citet{SL2010}, the process $\boldsymbol{Y}$ is a \Levy\ process. The expression of its characteristic function and that of its linear correlation coefficient are known in closed form and they are respectively given by

\begin{equation*}
	\phi_{\boldsymbol{Y}(t)\left(\boldsymbol{u}\right)} = \prod_{i=1}^{n}\phi_{I_{i}(t)}\left(u_{i}\mu_{i} + \frac{i}{2}\sigma_{i}^{2}u_{i}^{2}\right) \phi_{Z_a(t)}\left(\boldsymbol{u}^{T}\boldsymbol{\mu} + \frac{i}{2}\boldsymbol{u}^{T}\Sigma \boldsymbol{u}\right),
\end{equation*} 
where $\boldsymbol{\mu} = \left[\alpha_{1}\mu_{1},\dots,\alpha_{n}\mu_{n}\right]$ and
\[\Sigma =
\begin{bmatrix}
	\alpha_{1}\sigma_{1}^{2} & \sqrt{\alpha_{1}\alpha_{2}}\sigma_{1}\sigma_{2}\rho_{1,2} & \dots & \sqrt{\alpha_{1}\alpha_{n}}\sigma_{1}\sigma_{n}\rho_{1,n} &   \\
	\vdots & \ddots & \ddots & \vdots \\
	\alpha_{n}\sigma_{n}^{2} & \sqrt{\alpha_{n}\alpha_{1}}\sigma_{n}\sigma_{1}\rho_{n,1} & \dots & \sqrt{\alpha_{n}\alpha_{1}}\sigma_{n}\sigma_{1}\rho_{n,1} 
\end{bmatrix},
\]
and
\begin{equation*}
	\rho_{Y_{i}\left(t\right),Y_{j}\left(t\right)} = \frac{\mu_{i}\mu_{j}\alpha_{i}\alpha_{j}Var\left[Z_a\left(t\right)\right] +  \rho\sigma_{i}\sigma_{j}\sqrt{\alpha_{i}\alpha_{j}}\E\left[Z_a\left(t\right)\right]}{\sqrt{Var\left[Y_{i}\left(t\right)\right]Var\left[Y_{j}\left(t\right)\right]}}.
\end{equation*}
In this case we observe that, even if the drift $\mu_{j}$ vanishes the linear correlation coefficient can still be different from zero. Furthermore, if the correlation $\rho=0$ we obtain the same expression of the original model proposed by \citet{Semeraro2008}. 

\subsection{Ballotta-Bonfiglioli's approach}
Alternatively to the previous approaches, the dependence can be also introduced at the level of subordinated Brownian motions as proposed by \citet{BB2013} (hereafter labeled BB) . This approach is intuitive, but some attention is required in order to guarantee that the sum of two subordinated Brownian motions is still a subordinated Brownian motion of the same family. In order to ensure that, one has to impose some convolutions conditions. Unfortunately, in some case, they can be difficult to handle, especially from a numerical point of view.

\subsubsection{Convolution conditions}
In this section first we investigate the convolution conditions which guarantee that a linear combination of two random variables with a \VGpp\ keeps the same law. Accordingly, we use such conditions in order to build a multidimensional version of the \VGpp\ process.

Consider three independent subordinators $G_{Y},G_{X}$ and $G_{Z}$ with $\Gamma^{++}$ law such that at time $t\ge 0$:
\begin{align*}
	& G_{Y}(t)  \sim \Gamma^{++}\left(a,A_{y},B_{y}\right), \quad A_{y},B_{y}>0,\\
	& G_{X}(t)  \sim \Gamma^{++}\left(a,A_{x},B_{x}\right), \quad A_{x},B_{x}>0,\\
	& G_{Z}(t)  \sim \Gamma^{++}\left(a,A_{z},B_{z}\right), \quad A_{z},B_{z}>0,.
\end{align*} 
and three independent Brownian motions $W_{x},W_{y}$ and $W_{z}$. We define then, the processes $X,Y$ and $Z$ as follows
\begin{align*}
	Y(t) &= \theta G_{Y}(t) + \sigma W_{y}\left(G_{Y}(t)\right), \quad \theta \in \R,\; \sigma >0,\\
	X(t) &= \beta  G_{X}(t) + \gamma W_{x}\left(G_{X}(t)\right), \quad \beta \in \R,\; \gamma >0,\\
	Z(t) &= \beta_{z} G_{Z}(t) + \gamma_{z}W_{z}\left(G_{Z}(t)\right), \quad \beta_{z} \in \R,\; \gamma_{z} >0.
\end{align*}
Finally, we define the process $\hat{Y}= X + a_{1}Z$, with $a_{1} \in \R$. The goal is to find some relations between the parameters such that $Y(t) \eqd \hat{Y}(t),\; \forall t \ge 0$. Following the approach of \citet{BB2013} and using \eqref{eqn:chf_Zsubordinator} we equate the characteristic exponents, obtaining

\begin{equation*}
	A_{y} \log\left(\frac{1-iya/B_{y}}{1-iy/B_{y}}\right) = A_{x} \log\left(\frac{1-ixa/B_{x}}{1-ix/B_{x}}\right) + A_{z} \log\left(\frac{1-iza/B_{z}}{1-iz/A_{z}}\right),
\end{equation*}
where
\begin{align*}
y &= u\theta + iu^{2}\frac{\sigma^{2}}{2},\\
x &= u\beta + iu^{2}\frac{\gamma^{2}}{2},\\
z &= u\beta_{z}a_{1} + iu^{2}a_{1}^{2}\frac{\gamma_{z}^{2}}{2}.
\end{align*}
The first condition we need to impose is $A_{y} = A_{x} + A_{z}$ which leads to
\begin{equation*}
	\begin{split}
	A_{y} \log\left(\frac{1-iya/B_{y}}{1-ix/B_{y}}\right) & = A_{y} \log\left(\frac{1-ixa/B_{x}}{1-ix/B_{x}}\right) - A_{z} \log\left(\frac{1-ixa/B_{x}}{1-ix/B_{x}}\right)\\ &+ A_{z} \log\left(\frac{1-iza/B_{z}}{1-iz/A_{z}}\right).
	\end{split}
\end{equation*}
The second one is
\begin{align*}
	\frac{y}{B_{y}} & = \frac{x}{B_{x}}, \\
	\frac{x}{B_{x}} & = a_{1}\frac{z}{B_{z}}.
\end{align*}
From these equations we obtain the following conditions
\begin{align*}
	\beta & = \frac{B_{x}}{B_{z}}a_{1}\beta_{z}, \qquad \gamma^{2} = \frac{B_{x}}{B_{z}}\gamma_{z}a_{1}^{2}, \\
	\theta &= \frac{B_{y}}{B_{x}}\gamma^{2} \beta,\qquad \sigma^{2} = \frac{B_{y}}{B_{x}}\gamma^{2},
\end{align*}
and therefore
\begin{equation*}
	\E\left[Y(t)\right]= \frac{A_{y}}{B_{y}}\theta= \beta\frac{A_{x}}{B_{x}} + a_{1}\beta_{z}\frac{A_{z}}{B_{z}}.
\end{equation*}
Finally, it is easy to check that
\begin{equation*}
	Var\left[Y(t)\right] = \theta^{2}\left(1-a^2\right) \frac{A_{y}}{B_{y}^{2}} + \sigma^{2}\left(1-a^{2}\right) \frac{A_{y}}{B_{y}}.
\end{equation*}

\noindent Consequently, we have $Y(t) \eqd \tilde{Y}(t),\; \forall t \ge 0$ and a tool to construct multidimensional \Levy\ processes as proposed by \citet{BB2013}.

\subsubsection{A multidimensional \VGpp\ process}
According to the convolution conditions of the previous section we can introduce a multi-dimensional version of the \VGpp\ as follows
\begin{defn}[\VGpp\ Ballotta and Bonfiglioli's model]
	\label{def:BBmodelsd}
	Let $Z_a$ be a \sd\ subordinator and define the process \textcolor{black}{$\boldsymbol{Y} = \left\{\boldsymbol{Y}\left(t\right); t \ge 0\right\}$} as
	
	\begin{equation}
		\boldsymbol{Y}\left(t\right) = \left(Y_{1}\left(t\right),\dots,Y_{n}\left(t\right)\right) = \left(X_{1}\left(t\right) + a_{1}Z\left(t\right),\dots, X_{n}\left(t\right) + a_{n} Z\left(t\right)\right),
		\label{eqn:extendedModelBB}
	\end{equation}
	where
	
	\begin{itemize}
		\item \textcolor{black}{$X_{j}=\left\{X_{j}\left(t\right); t \ge 0\right\}$} is a subordinated Borwnian motion with parameters $\left(\beta_{j},\gamma_{j},\nu_{j}\right)$, $j=1,\dots,n$, where $\beta_{j} \in \mathbb{R}$ is the drift, $\gamma_{j}\in \mathbb{R}^{+}$ is the diffusion and $\nu_{j} \in \mathbb{R}^{+}$ is the variance of the subordinator. Let be $G_{X_{j}} = \left\{G_{X_{j}}\left(t\right); t \ge 0\right\}$ the subordinator of $X_{j}$ and let be $G_{X_{j}}$ be mutually independent for $j=1,\dots,n$. We define
		\begin{equation*}
			X_{j}\left(t\right) = \beta_{j}G_{X_{j}}\left(t\right) + \gamma_{j} W_{x_{j}}\left(G_{X_{j}}\left(t\right)\right),\quad j=1,\dots,n.
		\end{equation*}
		\item Define $Z = \left\{Z\left(t\right); t \ge 0\right\}$:
		\begin{align}
			Z\left(t\right) & = \beta_{z}Z_a\left(t\right) + \gamma_{z}W_{z}\left(Z_a\left(t\right)\right), \nonumber \\
			\end{align}
		where $W_{z} = \left\{W_{z}\left(t\right); t \ge 0\right\}$ is a \BM\ and $\beta_{z}\in \mathbb{R}$ and $\gamma_{z} \in \mathbb{R}^{+}$.
		\item \begin{align*}
        	 G_{X_{j}}(t)  & \sim \Gamma^{++}\left(a,A_{x_{j}}t,B_{x_{j}}\right),\\
        	 Z_a(t)  & \sim \Gamma^{++}\left(a,A_{z}t,B_{z}\right).
        \end{align*} 
		\item Let $a_{1},\dots,a_{n} \in \mathbb{R}$.
	\end{itemize}
\end{defn}
The resulting process is such that $Y_{j} \eqd \theta_{j}G_{Y_{j}}(t) + \sigma_{j} W_{y}(G_{Y_{j}}(t)) $ where $G_{Y_{j}}(t) \sim \Gamma^{++}\left(a,A_{y_{j}}t,B_{y_{j}}\right)$. In financial modeling it is customary to require that $\E\left[G_{Y_{j}}(t)\right]=t$ which can be obtained by imposing $B_{y_{j}} = \left(1-a\right)A_{y_{j}}$.

After straightforward computations one can show that the characteristic function of the process $\boldsymbol{Y}$ at time $t$ is given by

\begin{equation*}
	\phi_{\boldsymbol{Y}(t)}\left(\boldsymbol{u}\right) = \prod_{k=1}^{n} \phi_{G_{k}(t)}\left(\beta_{k}u_{k} + \frac{i}{2}u_{k}^{2}\gamma_{k}^{2}\right) \cdot \phi_{Z_a(t)}\left(\sum_{k=1}^{n}a_{k}u_{k}\beta_{z} + \frac{i}{2}\left(\sum_{k=1}^{n}a_{k}u_{k}\gamma_{z}\right)^{2}\right).
\end{equation*}
Finally, the linear correlation coefficient at time $t$ has the following expression

\begin{equation*}
	\rho_{Y_{i}(t),Y_{j}(t)} = \frac{a_{i}a_{j}\left(\beta_{z}^{2}\left(1-a^{2}\right)A_{z}/B_{z}^{2} + \left(1-a\right)\gamma_{z}^{2}A_{z}/B_{z}\right)}{\sqrt{\theta_{i}^{2}\left(1-a^2\right) \frac{A_{y_{i}}}{B_{y_{i}}^{2}} + \sigma_{i}^{2}\left(1-a^{2}\right) \frac{A_{y_{i}}}{B_{y_{i}}}}\cdot \sqrt{\theta_{j}^{2}\left(1-a^2\right) \frac{A_{x_{j}}}{B_{x_{j}}^{2}} + \sigma_{j}^{2}\left(1-a^{2}\right) \frac{A_{y_{j}}}{B_{y_{j}}}}}.
\end{equation*}

\section{Numerical Experiment}
\label{sec:NumericalResults}
In this section we apply the models we have studied so far to real data and  investigate their performance. We focus on the multidimensional versions of the VG and \VGpp\ process presented in Section \ref{sec:MargrabeVG} and \ref{sec:MargrabeVGpp} and on those discussed in Section \ref{sec:MultidimensionalModeling}. For each of them we have calibrated the model on market data and priced exchange options using different numerical techniques.
\par In order to calibrate our models we need the quoted prices of the derivatives contracts written on each of the two forward products and their joint historical time series. The data-set\footnote{Data Source: www.eex.com.} we rely upon consists of
\begin{itemize}
	\item Forward quotations from 4 January 2021 to 19 May 2022 of calendar 2023 power forward. A forward calendar 2023 contract is a contract to buy or sell a specific volume of energy in MWh at fixed price for all the hours of 2023. Calendar power forward in Germany and France are denoted with DEBY and F7BY, respectively.
	\item Quotations of European call options on power forward 2023 in Germany and France with settlement date 19 May 2022. We use strikes in a range of $\pm 15\, [EUR/MWh]$ around the settlement price of the forward contract, i.e. we excluded deep ITM and OTM options.
	\item We assume a risk-free rate $r=0.015$.
	\item The estimated historical correlation between markets log-returns is $\rho_{mkt} = 0.96$.
\end{itemize} 

We have adopted the same two-steps calibration procedure of \citet{SL2010}: to this end, it is worthwhile noticing that the marginal distributions do not depend on the parameters required to model the dependence structure. The vector of the marginal parameters $\Theta^{*}$ is obtained solving the following optimization problem:
\begin{equation}
	\Theta^{*} = \argmin_{\Theta} \sum_{i=1}^{n} \left(C_{i}^{\Theta}\left(K,T\right) - C_{i}\right)^{2},
	\label{eqn:minimizationproblem_margins}
\end{equation}
where $C_{i}, i=1, \dots, n$ are the values of $n$ quoted vanilla products and $C_{i}^{\Theta}\left(K,T\right), i=1, \dots, n$ are the relative model prices.
Once  $\Theta^{*}$ is found we have to calibrate the remaining parameters for the dependence structure. Generally derivative contracts written on multiple underlying assets are not very liquid and market quotes are rarely available therefore, the vector $\boldsymbol{\eta}^{*}$, that encompasses the dependence parameters, has been estimated fitting the correlation matrix on historical data. The expression of the theoretical correlation matrix for the extension of \citet{SL2010} and \citet{BB2013} in the \VGpp\ framework has been derived in Section \ref{sec:MultidimensionalModeling}: the parameters have been chosen so that the correlation is fitted accordingly to the market.\\
\par  For the first calibration step we have combined the NLLS approach with the FFT method proposed by \citet{Carr99} (the version proposed by \citet{lewis} returns similar results) whereas for the second one, we have used the plain NLLS method for the minimization of the distance between the theoretical and the observed correlation coefficient. The calibrated parameters are reported in Tables \ref{tbl:BlackScholes_parameters}, \ref{tbl:VG_parameters}, \ref{tbl:VG++_parameters}, \ref{tbl:VG++LS_parameters} and \ref{tbl:VG++BB_parameters} , whereas Table \ref{tbl:exchange_options_prices} reports the values of the exchange option computed with Monte Carlo, the FFT method and the closed formula expression, when available.

\begin{table}[!htb]
	\scriptsize
	
	\begin{minipage}{.35\linewidth}
		\centering
		\begin{tabular}[t]{cc}
		\toprule
		Parameter & Value \\ [0.5ex]
		\midrule
		$\sigma_{1}$ & 0.84  \\
		$\sigma_{2}$ & 0.91  \\
		$\rho$ & 0.96 \\
				\bottomrule
		 &  \\
		 &  \\ 
		 &  \\ 
		 &  \\ [1ex]
	\end{tabular}
		\vspace{0.15cm}
		\caption{\emph{BS}}
		\label{tbl:BlackScholes_parameters}
	\end{minipage}%
	\begin{minipage}{.35\linewidth}
		\centering
		\begin{tabular}[t]{cc}
			\toprule
			Parameter & Value \\ [0.5ex]
			\midrule
			$\mu_{1}$ & -0.27  \\
			$\mu_{2}$ & -0.24  \\
			$\sigma_{1}$ & 0.98  \\
			$\sigma_{2}$ & 0.92  \\
			$\alpha$ & 2.04 \\
			$\rho$ & 0.96 \\
			\bottomrule
			 &  \\ [1ex]
		\end{tabular}
		\vspace{0.15cm}
		\caption{\emph{VG}}
		\label{tbl:VG_parameters}
	\end{minipage}%
	\begin{minipage}{.35\linewidth}
		\centering
		\smallskip
		\begin{tabular}[t]{cc}
			\toprule
			Parameter & Value \\ [0.5ex]
			\midrule
			$\mu_{1}$ & -0.28  \\
			$\mu_{2}$ & -0.26  \\
			$\sigma_{1}$ & 0.98  \\
			$\sigma_{2}$ & 0.92  \\
			$\alpha$ & 2.43 \\
			$a$ & 0.04 \\
			$\rho$ & 0.96 \\ [1ex]
			\bottomrule
		\end{tabular}
		\vspace{0.15cm}
		\caption{\emph{\VGpp}}
		\label{tbl:VG++_parameters}
	\end{minipage}%
\vspace{1cm}
	\begin{minipage}{.4\linewidth}
	\centering
	\smallskip
	\begin{tabular}[t]{cc}
		\toprule
		Parameter & Value \\ [0.5ex]
		\midrule
		$\mu_{1}$ & -0.47  \\
		$\mu_{2}$ & -0.35  \\
		$\sigma_{1}$ & 1.06  \\
		$\sigma_{2}$ & 0.96  \\
		$\alpha_{1}$ & 1.43 \\
		$\alpha_{2}$ & 1.64 \\
		$a$ & 0.01 \\
		$A$ & 1.43 \\
		$B$ & 1224.83 \\
		$\rho$ & 0.99 \\
		[1ex]
		\bottomrule
	\end{tabular}
	\vspace{0.15cm}
	\caption{\emph{LS-\VGpp }}
	\label{tbl:VG++LS_parameters}
\end{minipage}%
\begin{minipage}{0.7\linewidth}
		\centering
		\smallskip
		\begin{tabular}[t]{cccc}
			\toprule
			Parameter & Value & Parameter & Value\\ [0.5ex]
			\midrule
			$\mu_{1}$ & -0.38  & $\mu_{2}$ & -0.33  \\
			$\sigma_{1}$ & 1.025  & $\sigma_{2}$ & 0.96  \\
			$A_{y_{1}}$ & 1.5839 & $A_{y_{2}}$ & 1.4439 \\
			$A_{x_{1}}$ & 0.141 & $A_{x_{2}}$ & 0.0001 \\
			$B_{x_{1}}$ & 12.20 & $B_{x_{2}}$ & 0.006 \\
			$\beta_{1}$ & -2.95 & $\beta_{2}$ & -0.00014 \\
			$\gamma_{1}$ & 2.84 & $\gamma_{2}$ & 0.062 \\
			$a_{1}$ & 4.80 & $a_{2}$ & 4.74 \\
			$A_{z}$ & 1.4438 & $B_{z}$ & 62.77 \\
			$\beta_{z}$ & -3.15 & $\gamma_{z}$ & 1.34 \\
			$a$ & 0.001 & & \\ [1ex]
			\bottomrule
		\end{tabular}
		\vspace{0.15cm}
		\caption{\emph{BB-\VGpp}}
		\label{tbl:VG++BB_parameters}
	\end{minipage}
\end{table}

\begin{table}
	\centering
		\begin{tabular}[t]{cccc}
	\toprule
	Model & Monte Carlo & Exact & FFT\\ [0.5ex]
	\midrule
	BS & 80.04 & 79.92  & 79.91 \\
	VG & 81.69 & 81.68  & 81.65 \\
	\VGpp & 81.62 & 81.28  & 81.15 \\
	\VGpp\ LS & 85.18 & -  & 85.19 \\
	\VGpp\ BB & 83.15 & -  & 83.29 \\ [1ex]
	\bottomrule
\end{tabular}
\caption{\emph{Exchange option prices}}
\label{tbl:exchange_options_prices}
\end{table}

From Table \ref{tbl:exchange_options_prices} we can observe that all models return very similar prices and that all of them can replicate the observed market log-returns correlation level with the exception of the LS-\VGpp\ model which attains a slightly lower level, $\rho_{mod} = 0.93$. Since the value of the spread option is very sensitive to the correlation, in order to avoid mispricing, one should avoid selecting a model which does not fit it properly.
\par We remark that we have constructed the VG and \VGpp\ models with a single subordinator whereas, in the extended models LS-\VGpp\ and BB-\VGpp\ we have introduced different dependent subordinators. We observe that the option prices obtained with the different constructions are very similar. This fact is somehow intuitive from an economical point of view. Indeed, the German and French power futures markets are extremely intertwined  because of the physical electrical grid connections between the two related countries. We could say that the stochastic time \enquote{runs in the same way} in both markets and hence, using a single subordinator for price modeling is not a restrictive choice. On the other hand, there may be other market situations in which multi-subordination might be the right approach (see \citet{SL2010} and \citet{BB2013}). In general, when different models lead to comparable results, it is reasonable to select the simpler one and avoid unnecessary complexity.
\par In Section \ref{sec:VGpp_process} we have briefly mentioned that the parameter $a\in\left(0,1\right)$ can be interpreted as an index of liquidity. Here we observe that $a$, and hence the probability of having a zero increment in the underlying price over the time interval $\Delta t$, is very close to zero in all cases. During the previous years the options power market has lacked of liquidity in many European countries but recently such derivative instruments have become more and more popular leading to an increase in the number of exchanged trades.
\par Possible limitations in the construction of the multidimensional version of the \VGpp\ processes might arise if we consider markets with different level of liquidities. Indeed, in order to use the summing property of the $\Gamma^{++}$ law, the value of  $a\in \left(0,1\right)$ must be the same for all the process we consider. For this reason, such a construction is reasonable if the considered markets have approximately the same level of liquidity. This fact might seem restrictive but, in the most important European power options markets, almost the same level of liquidity is often observed and hence, the proposed approach results reasonable.

\par In this section we have focused only on exchange options, namely spread options where $K$ is equal to zero. If the strike is not zero, no closed form expression is know, even in the most simple case of the BS model. Nevertheless, one can use the approximate formula proposed by \citet{Kirk1995} or the numerical methods proposed by \citet{hurd2009}, \citet{caldanafusai2016} or \citet{PellegrinoSabino_1}. 

\section{Conclusions}
\label{sec:conclusions}

In this article we have firstly discussed the problem of spread option pricing when the dynamic of the log-prices follows a VG or a \VGpp\ process. Focusing on exchange options, we have derived closed form pricing formulas when all the involved Brownian motions are subordinated by a single subordinator. In particular, for the multidimensional version of the \VGpp\ process of Section \ref{sec:MargrabeVGpp} we have derived an integral free formula. Accordingly, we have investigated the numerical performance of the derived formulas, comparing the results with the ones obtained by Monte Carlo and Fourier techniques. The explicit option pricing formulas are accurate and efficient, leading to computational benefits with respect to the Monte Carlo method, especially in the variance gamma case.

As a second contribution, we have proposed more general versions of the multidimensional \VGpp\ model. Instead of using a common subordinator, these versions have been obtained by time changing each Brownian motion with different dependent subordinator relying upon the summing and scaling properties of the $\Gamma^{++}$ distribution.
This new \Levy\ framework leads to tractable models, both from a theoretical and numerical point of view. For instance,  the relative expressions of the characteristic function and of the linear correlation coefficient can be easily derived. Such results are instrumental for the model calibration and for the  option pricing based on FFT techniques.
\par Finally, we have applied all these models to the German and French power futures market. We have calibrated them on vanilla contracts and on historical futures quotations and accordingly have priced exchange options. Under these assumptions we have shown that the models with a single subordinator are sufficiently accurate. This fact has a clear economic interpretation since the German and French electrical grids are strongly connected and therefore, one can assume that the business time modeled by the subordinator \enquote{flows equally fast} in both markets. Furthermore, we have observed that the self-decomposability parameter $a$, that is related to the market liquidity, is very small and reflects the fact that the option market is sufficiently liquid, as has been recently observed also from an empirical point of view.
In this work we have mainly focused on the pricing of exchange options nevertheless, all numerical methods could be easily adapted  for the pricing of spread options. On the other hand, the investigation of approximate formulas  for spread options, for instance, in Kirk's spirit, under the VG and the \VGpp\ models  could be of some interest and this will be the subject of a future research.


\bibliographystyle{plainnat}
\bibliography{library.bib}

\end{document}